\documentclass[10pt]{article}
\usepackage[dvips]{graphicx}

\usepackage{epsfig}
\usepackage{graphicx}
\usepackage{indentfirst}
\usepackage{amsmath}
\usepackage{amsfonts}
\usepackage{amssymb}

\begin{document}

\begin{center}
{\Large\bf  Finite-time future singularities models in $f(T)$ gravity and the effects of viscosity\\}
\medskip

  M. R. Setare$^{a}$\footnote{e-mail:
rezakord@ipm.ir} and   M. J. S. Houndjo$^{b,c}$\footnote{e-mail:
sthoundjo@yahoo.fr}\\

$^{a}$ { \it Department of Science, Payame Noor University, Bijar, Iran} 
\\
 
$^{b}$ { \it Departamento de Ci\^{e}ncias Naturais - CEUNES \\
Universidade Federal do Esp\'irito Santo\\
CEP 29933-415 - S\~ao Mateus - ES, Brazil}\\
$^{c}$ {\it Institut de Math\'{e}matiques et de Sciences Physiques (IMSP)}\\
 {\it 01 BP 613 Porto-Novo, B\'{e}nin}
\medskip\\   
\end{center}

\begin{abstract}
We investigate models of future finite-time singularities in $f(T)$ theory, where $T$ is the torsion scalar. The algebraic function $f(T)$ is put as the teleparallel term $T$ plus an arbitrary function $g(T)$. A suitable expression of the Hubble parameter is assumed and constraints are imposed in order to provide an expanding universe. Two parameters $\beta$ and $H_s$ that appear in the Hubble parameter are relevant in specifying the types of singularities. Differential equations of $g(T)$ are established and solved, leading to the algebraic $f(T)$ models for each type of future finite time singularity. Moreover, we take into account the viscosity in the fluid and discuss three interesting cases: constant viscosity, viscosity proportional to $\sqrt{-T}$ and the general one where the viscosity is proportional to $(-T)^{n/2}$, where $n$ is a natural number. We see that for the first and second cases, in general, the singularities are robust against the viscous fluid, while for the general case, the Big Rip and the Big Freeze can be avoided from the effects of the viscosity for some values of $n$.
\end{abstract}

Pacs numbers:

\section{Introduction}
Recent observations of type Ia supernova (SNIa) and WMAP \cite{1,2}
indicate that our universe is currently undergoing an accelerating
expansion, which confront the fundamental theories with great
challenges and also make the researches on this problem a major
endeavour in modern astrophysics and cosmology. Missing energy
density - with negative pressure - responsible for this expansion
has been dubbed dark energy. Wide range of scenarios have been
proposed to explain this acceleration while most of them can not
explain all the features of universe or they have so many parameters
that makes them difficult to fit. In this direction we can consider
theories of modified gravity \cite{3}, or field models of dark
energy. The field models that have been discussed widely in the
literature consider a cosmological constant \cite{4}, a canonical
scalar field (quintessence)\cite{5}, a phantom field, that is a
scalar field with a negative sign of the kinetic term \cite{6,7}, or
the combination of quintessence and phantom in a unified model named
quintom \cite{8}. In the other hand modified models of gravity
provides the natural gravitational alternative for dark energy
\cite{9}. Moreover, modified gravity present natural unification of
the early time inflation and late-time acceleration thanks to
different role of gravitational terms relevant at small and at large
curvature. Also modified gravity may naturally describe the
transition from non-phantom phase to phantom one without necessity
to introduce the exotic matter. Among these theories, scalar-tensor
theories \cite{10}, $f(R)$ gravity \cite{11}, DGP braneworld gravity
\cite{12} and string-inspired theories \cite{13} are studied
extensively. Recently a theory of $f(T)$ gravity has been received
attention. Models based on modified teleparallel gravity were
presented, in one hand, as an alternative to inflationary models
\cite{14,15}, and on the other hand, as an alternative to dark
energy models \cite{16}. New spherically symmetric solutions of black holes and wormholes are obtained with a constant torsion and for the cases for which the radial pressure is proportional to a real constant, to some algebraic functions $f(T)$ and their derivatives, or vanishing identically \cite{daouda1}. In the same way, an algebraic function $f(T)$ is obtained through the reconstruction method for two cases and the study of a polytropic model for the stellar structure is developed \cite{daouda2}. Moreover, $f(T)$ gravity is reconstructed according to holographic dark energy is explicitly presented in \cite{daouda3} and latter an anisotropic fluid for a set of non-diagonal tetrads in $f(T)$ gravity explored generating various classes of new black hole and wormhole solutions \cite{daouda4}. Also, many works have been done in order to check whether $f(T)$ gravity can present results consistent with the many advances in cosmology and astrophysics \cite{bamba18}. Recently, Bamba et al investigate the reconstruction of power law model, exponential model and logarithmic model, able to reproduce some of the future finite time singularities and also discuss the thermodynamics near these singularities \cite{odin} (For other works about future finite time singularities, see \cite{suggestedsergei}). { \bf Also, future singularities with the presence of a viscous fluid as well as other interesting properties, have been already studied (in the context of General Relativity and f(R) gravity)\cite{diegosg}}. 
\par
In the present paper we investigate the $f(T)$ gravity  models that are able to reproduce the four types of finite time future singularities from a suitable choice of Hubble parameter. A parameter $\beta$ in the expression of the Hubble parameter plays an important role in specifying these singularities. The algebraic function $f(T)$ is assumed as the sum of the teleparallel term $T$ and an algebraic function $g(T)$ with which all the task is done. According to some values of the parameter $\beta$, differential equation of $g(T)$ are established and solved in some ways.  The algebraic function $f(T)$ for each type of singularity is obtained from each expression of $g(T)$.\par 
On the other hand, we notify that the presence of finite-time future singularities may cause serious problems in the black holes or stellar astrophysics \cite{gorbu9}. A way to probe the possible avoidance of these singularities is considering that the fluid possesses viscosity. This is the second purpose of this work. As previously mentioned,  the models that lead to future finite time singularities have been reconstructed considering a non viscous fluid. Thus, we introduce the viscosity and investigate its effects on the  singularities. We see that when the constant viscosity or the viscosity proportional to $\sqrt{-T}$ is considered, in general, the singularities are robust against the viscosity. However, when the viscosity is proportional to $(-T)^{n/2}$, for some values of the parameter $n$, the viscosity may cure the Big Rip and the Big Freeze.

\par
The paper is organized as follows. In Sec. $2$, the $f(T)$ gravity formalism and the field equations are presented. The Sec. $3$ is devoted to the presentation of the Hubble parameter, the classification of future finite time singularities. Suitable scale factor coming from the Hubble parameter are presented according to the values of the parameter $\beta$ for obtaining the algebraic $f(T)$ function. The viscosity is introduced in the Sec. $4$ and its effect is investigated as the singularities are approached. The conclusion and perspectives are presented in Sec. $5$.

\section{$f(T)$ gravity and field equations }
Let us define the notation of the Latin subscript as those related to the tetrad fields, and the Greek one related to the spacetime coordinates. For a general specetime metric, we can define the line element as 
\begin{eqnarray}
ds^2=g_{\mu\nu}dx^\mu dx^\nu\,\,\,.
\end{eqnarray}
The projection of this line element can be described in the tangent space to the spacetime through the matrix called tetrad as follows:
\begin{eqnarray}
ds^2=g_{\mu\nu}dx^{\mu}dx^{\nu}= \eta_{ij}\theta^{i}\theta^{j}\,\,\,,\\
dx^\mu=e_{i}^{\,\,\,\mu}\theta^{i}\,\,\,, \theta^{i}=e^{i}_{\,\,\,\mu}dx^{\mu}\,\,\,,
\end{eqnarray}
where $\eta_{ij}$ is the metric on Minkowski's spacetime and $e_{i}^{\,\,\,\mu}e^{i}_{\,\,\,\nu}=\delta^{\mu}_{\nu}$, or $e_{i}^{\,\,\,\mu}e^{j}_{\,\,\,\mu}=\delta^{j}_{i}$.\par
The action for the theory of modified gravity based on a
modification of the teleparallel equivalent of General
Relativity, namely $f(T)$ theory of gravity, coupled with matter
$L_m$ is given by \cite{15,19,20,setareparticle}
\begin{equation}\label{set1}
S=\frac{1}{16\pi G}\int d^4x e \left[T+g(T)+L_m\right]_,
\end{equation}
where $e=det(e^i_{\,\,\mu})=\sqrt{-g}$. Here, $G$ is the gravitational constant and $c$ the speed of the light. From now, we will use the units $8\pi G=c=1$. The teleparallel Lagrangian
$T$ is defined as follows
\begin{equation}\label{set2}
T=S^{\:\:\:\mu \nu}_{\rho} T_{\:\:\:\mu \nu}^{\rho},
\end{equation}
where
\begin{eqnarray}
T_{\:\:\:\mu \nu}^{\rho}=e_i^{\;\;\rho}(\partial_{\mu}
e^i_{\;\;\nu}-\partial_{\nu} e^i_{\;\;\mu}),\label{set3}\\
S^{\:\:\:\mu \nu}_{\rho}=\frac{1}{2}(K^{\mu
\nu}_{\:\:\:\:\:\rho}+\delta^{\mu}_{\rho} T^{\theta
\nu}_{\;\;\:\:\:\theta}-\delta^{\nu}_{\rho} T^{\theta
\mu}_{\;\;\:\:\:\theta}),\label{set4}
\end{eqnarray}
and $K^{\mu \nu}_{\:\:\:\:\:\rho}$ is the contorsion tensor
\begin{eqnarray}
K^{\mu \nu}_{\:\:\:\:\:\rho}=-\frac{1}{2}(T^{\mu
\nu}_{\:\:\:\:\:\rho}-T^{\nu \mu}_{\:\:\:\:\:\rho}-T^{\:\:\:\mu
\nu}_{\rho}).\label{set5}
\end{eqnarray}
The field equations are obtained by varying the action with
respect to vierbein $e^i_{\mu}$ as follows
\begin {equation}\label{set6}
-e^{-1}\partial_{\mu}(e S^{\:\:\:\mu
\nu}_{i})(1+g_T)-e_i^{\:\lambda}T_{\:\:\:\mu
\lambda}^{\rho}S^{\:\:\:\nu \mu}_{\rho}g_T +S^{\:\:\:\mu
\nu}_{i}\partial_{\mu}(T)g_{TT}-\frac{1}{4}e_{\:i}^{\nu}
(T+g(T))=  e_i^{\:\rho}\mathcal{T}_{\rho}^{\:\:\nu},
\end{equation}
where $g_T=g'(T)$ and $g_{TT}=g''(T)$ and $\mathcal{T}$ the energy momentum tensor. Now, we take the usual
spatially-flat metric of Friedmann-Robertson-Walker (FRW)
universe, in agreement with observations
\begin {equation}\label{set7}
ds^{2}=dt^{2}-a(t)^{2}\sum^{3}_{i=1}(dx^{i})^{2},
\end{equation}
where $a(t)$ is the scale factor as a one-parameter function of
the cosmological time $t$.\par 
Let us assume first that the background is a non-viscous fluid. Using the Friedmann-Robertson-Walker metric
and the perfect fluid matter in the Lagrangian
(\ref{set1}) and the field equations (\ref{set6}), one obtains
\begin {eqnarray}
T&=&-6H^2\,\,\,,\label{set8}\\
3H^2&=& \rho_{eff}\,\,\,,\label{set9}\\
-3H^2-2\dot{H}&=& p_{eff}\,\,,\label{set10}
\end{eqnarray}
where $\rho_{eff}$ and $p_{eff}$ denote respectively the effective energy density and pressure of the universe and defined by
\begin{eqnarray}
\rho_{eff}&=&\rho-\frac{1}{2}g-6H^2g_T\,\,\,,\label{set11}\\
p_{eff}&=& p+\frac{1}{2}g+2\left(3H^2+\dot{H}\right)g_T-24\dot{H}H^2g_{TT}\,\,\,,\label{set12}
\end{eqnarray}
where $H$ is the Hubble parameter and defined by
$H=\dot{a}/a$. Using (\ref{set11}) and (\ref{set12}), and combining (\ref{set9}) and (\ref{set10}), one gets
\begin{eqnarray}\label{set13}
2Tg_{TT}+g_T+\frac{\rho(1+\omega)}{2\dot{H}}+1=0\,\,\,,
\end{eqnarray}
where we used the barotropic equation of state $p=\omega \rho$. Then, for a given scale factor corresponding to a future finite time singularity, the action may explicitly be reconstructed by solving the differential equation (\ref{set13}).
\section{Future finite time singularities}
We propose to find in $f(T)$ gravity, models that reproduce the four  types of finite time future singularities from the Hubble parameter \cite{proposergei}
\begin{eqnarray}\label{set14}
H=h\left(t_s-t\right)^{-\beta}+\mbox{C}\,\,\,,
\end{eqnarray}
where $h$,  $C$ and $t_s$ are positive constants and $t<t_s$. These
constraints are imposed to the parameter for providing an
expanding universe. The parameter $\beta$ can be a positive constant or a negative
non-integer number. Then, as the singularity time $t_s$ is
approached, $H$ or some of its derivatives and therefore, the
torsion, diverge. $C$ is essentially relevant near the singularity
only when $\beta<0$ (where we denote it as $C=H_s$, the Hubble
parameter at the singularity time), and then, we can assume $C=0$
when $\beta>0$.\par The finite-time singularities are classified in
the following way \cite{gorbu8,step4}\par $\bullet$ Type I
(Big Rip): for $t\rightarrow t_s$, $a\rightarrow\infty$,
$\rho_{eff}\rightarrow \infty$ and $|p_{eff}|\rightarrow \infty$ at $t=t_s$. This
corresponds to $\beta =1$ and $\beta>1$.\par $\bullet$ Type II
(Sudden): for $t\rightarrow t_s$, $a\rightarrow a_s$,
$\rho_{eff}\rightarrow \rho_s$ and $|p_{eff}|\rightarrow \infty$. It
corresponds to $-1<\beta< 0$.\par $\bullet$ Type III (Big Freeze): for
$t\rightarrow t_s$,  $a\rightarrow a_s$,  $\rho_{eff}\rightarrow
\infty$ and $|p_{eff}|\rightarrow \infty$. This corresponds to
$0<\beta<1$. \par $\bullet$ Type IV (Big Brake): for $t\rightarrow t_s$,
$a\rightarrow a_0$, $\rho_{eff}\rightarrow 0$, $p_{eff}\rightarrow
0$ and higher derivatives of $H$ diverge,  which corresponds to the
case $\beta<-1$ but $\beta$ is not any integer number.\par Let us
now investigate the $f(T)$ gravity models for which the finite time
future singularities could occur, when (\ref{set14}) is assumed.
\subsection{Big Rip singularity models without viscosity}
This sort of singularity may appear for $\beta=1$ and $\beta>1$. Let us treat the cases separately.
\subsubsection*{The case $\beta=1$}
In this case, the corresponding scale factor can be written as
\begin{eqnarray}\label{set15}
a(t)=a_0\left(t_s-t\right)^{-h}\,\,\,,
\end{eqnarray}
and then, the first derivative of the Hubble parameter reads
\begin{eqnarray}\label{set16}
\dot{H}=h\left(t_s-t\right)^{-2}\,\,\,.
\end{eqnarray}
Hence,  Eq. (\ref{set13}) becomes
\begin{eqnarray}\label{set17}
2Tg_{TT}+g_{T}+\frac{\rho_0(1+\omega)\left(t_s-t\right)^{3h(1+\omega)+2}}{2ha_0^{3(1+\omega)}}+1=0\,\,.
\end{eqnarray}
In the other hand, using (\ref{set8}), with the scale factor (\ref{set15}), one can write
\begin{eqnarray}\label{set18}
t_s-t=\left(-\frac{T}{6h^2}\right)^{-1/2}\,\,\,.
\end{eqnarray}
Thus, Eq. (\ref{set17}) takes a new form as
\begin{eqnarray}\label{set19}
2Tg_{TT}+g_T+\frac{\rho_0(1+\omega)}{2ha_0^{3(1+\omega)}}\left(-\frac{T}{6h^2}\right)^{-\frac{3}{2}h(1+\omega)-1}+1=0\,\,\,.
\end{eqnarray}
The general solution of (\ref{set19}) reads
\begin{eqnarray}\label{set20}
g(T)=\frac{-1}{1+2B}\left[1+2B+\frac{A}{1+B}\left(-\frac{T}{6h^2}\right)^B+2\left(1+2B\right)C_1\sqrt{-T}\right]T+C_2\,\,\,,
\end{eqnarray}
where $C_1$ and $C_2$ are integration constants, and $A$ and $B$ defined respectively as
\begin{eqnarray}\label{set21}
A=\frac{\rho_0(1+\omega)}{2ha_0^{3(1+\omega)}}\,\,,\quad\quad B=-\frac{3}{2}h(1+\omega)-1\,\,.
\end{eqnarray}
The corresponding algebraic function $f(T)$ reads
\begin{eqnarray}\label{set22}
f(T)=-\frac{AT}{(1+2B)(1+B)}\left(-\frac{T}{6h^2}\right)^B-2C_1T\sqrt{-T}+C_2\,\,\,.
\end{eqnarray}
Initial condition may be applied for finding the respective values of the constants $C_1$ and $C_2$. We can follow the same process as in \cite{daouda4}. We assume that at the early time, that we denote $t_0$, the corresponding value (the initial value) of the torsion scalar is $T_0$ such that
\begin{eqnarray}\label{set23}
\left(\frac{dT}{dt}\right)_{t=t_0}=-12h^2\left(-\frac{T_0}{6h^2}\right)^{3/2}\,\,\,.
\end{eqnarray}
The initial conditions imposed to the functions $f$ read
\begin{eqnarray}\label{set24}
\left(f\right)_{t=t_0}=T_0\,\,,\quad \quad \left(\frac{df}{dt}\right)_{t=t_0}=\left(\frac{dT}{dt}\right)_{t=t_0}\,\,\,.
\end{eqnarray}
Making use of these initial conditions, (\ref{set24}), one gets
\begin{eqnarray}\label{set25}
C_1=-\frac{1}{3\sqrt{-T_0}}\left[1+\frac{A}{1+2B}\left(-\frac{T_0}{6h^2}\right)^B\right]\,\,,\quad C_2=\frac{2T_0}{3}\left[1+\frac{A(1-2B)}{2(1+B)(1+2B)}\left(-\frac{T_0}{6h^2}\right)^B\right]\,\,\,\,.
\end{eqnarray}
Then, the algebraic function (\ref{set22}), with the constants (\ref{set25}), leads to the Big Rip when the fluid is free of viscosity.
\subsubsection*{The case $\beta>1$}
In this case, the corresponding expression of the scale factor is
\begin{eqnarray}\label{set26}
a(t)=a_0e^{\frac{h(t_s-t)^{1-\beta}}{\beta-1}}\,\,,
\end{eqnarray}
and the first derivative of the Hubble parameter reads
\begin{eqnarray}\label{set27}
\dot{H}=\beta h\left(t_s-t\right)^{-\beta-1}\,\,.
\end{eqnarray}
Thus, the Eq. (\ref{set13}) becomes
\begin{eqnarray}\label{set28}
2Tg_{TT}+g_T+\frac{\rho_0(1+\omega)(t_s-t)^{\beta+1}}{2\beta h a_0^{3(1+\omega)}}e^{-\frac{3h(1+\omega)(t_s-t)^{1-\beta}}{\beta-1}}+1=0\,\,\,.
\end{eqnarray}
Using Eq. (\ref{set8}) and the scale factor (\ref{set26}), one gets
\begin{eqnarray}\label{set29}
t_s-t=\left(-\frac{T}{6h^2}\right)^{-\frac{1}{2\beta}}\,\,\,,
\end{eqnarray}
and the differential equation (\ref{set28}) can be written as
\begin{eqnarray}\label{set30}
2Tg_{TT}+g_T+\frac{\rho_0(1+\omega)}{2\beta h a_0^{3(1+\omega)}}\left(-\frac{T}{6h^2}\right)^{-\frac{\beta+1}{2\beta}}e^{-\frac{3h(1+\omega)\left(-\frac{T}{6h^2}\right)^{\frac{\beta-1}{2\beta}}}{\beta-1}}+1=0\,\,\,.
\end{eqnarray}
Note that this equation is more complicated and cannot be solved trivially. However, it can be solved as the singularity is approached, i.e., when $t\rightarrow t_s$. Then, as the singularity is approached, the trace diverges ($T\rightarrow -\infty$), and (\ref{set30}) becomes
\begin{eqnarray}\label{set31}
2Tg_{TT}+g_T+1=0\,\,\,\,,
\end{eqnarray}
whose solution is
\begin{eqnarray}\label{set32}
g(T)=-T+2C_3\sqrt{-T}+C_4\,\,\,,
\end{eqnarray}
where $C_3$ and $C_4$ are integration constants.
Consequently, $f(T)$ is written as
\begin{eqnarray}\label{set33}
f(T)=2C_3\sqrt{-T}+C_4\,\,\,.
\end{eqnarray}
In this case, the initial value of the first derivative of the torsion with respect to the cosmic time reads
\begin{eqnarray}\label{set34}
\left(\frac{dT}{dt}\right)_{t=t_0}=-12\beta h^2\left(-\frac{T_0}{6h^2}\right)^{\frac{2\beta+1}{2\beta}}\,\,\,.
\end{eqnarray}
Making use of (\ref{set34}) and the initial conditions, (\ref{set24}),  one gets
\begin{eqnarray}\label{set35}
C_3=-\sqrt{-T_0}\,\,\,\,,\quad\quad C_4=-T_0\,\,\,.
\end{eqnarray}
Then the corresponding algebraic $f(T)$ is
\begin{eqnarray}\label{set36}
f(T)= -2\sqrt{T_0T}-T_0\,\,\,.
\end{eqnarray}
This model also leads to the Big Rip in a universe dominated by a non-viscous fluid.
\subsection{The Big Freeze models without viscosity}
The Big Freeze appears for $0<\beta<1$, and the corresponding scale factor is the same as in (\ref{set26}), which leads to (\ref{set30}). However, as we are dealing with the singularity of type III, Eq. (\ref{set30}) becomes
\begin{eqnarray}\label{set37}
2Tg_{TT}+g_T+\frac{\rho_0(1+\omega)}{2\beta ha_0^{3(1+\omega)}}\left(-\frac{T}{6h^2}\right)^{-\frac{\beta+1}{2\beta}}+1=0\,\,\,,
\end{eqnarray}
whose general solution is
\begin{eqnarray}\label{set38}
g(T)=-T+\frac{\beta\rho_0(1+\omega)}{h(\beta-1)a_0^{3(1+\omega)}}T\left(-\frac{T}{6h^2}\right)^{-\frac{\beta+1}{2\beta}}+2C_5\sqrt{-T}+C_6\,\,\,,
\end{eqnarray}
where $C_5$ and $C_6$ are integration constants. Then, $f(T)$ is written as
\begin{eqnarray}\label{set39}
f(T)= \frac{\beta\rho_0(1+\omega)}{h(\beta-1)a_0^{3(1+\omega)}}T\left(-\frac{T}{6h^2}\right)^{-\frac{\beta+1}{2\beta}}+2C_5\sqrt{-T}+C_6\,\,\,.
\end{eqnarray}
By using (\ref{set23}) and the initial conditions (\ref{set24}), the constant $C_5$ and $C_6$ are determined as 
\begin{eqnarray}\label{set40}
C_5=\frac{\rho_0(1+\omega)\sqrt{6}}{2a_0^{3(1+\omega)}}\left(-\frac{T_0}{6h^2}\right)^{-\frac{1}{2\beta}}\,\,\,,\quad C_6=T_0\left[1+\left(\frac{\beta\rho_0(1+\omega)}{h(\beta-1)a_0^{3(1+\omega)}}-\frac{\beta+1}{\beta}\right)\left(-\frac{T_0}{6h^2}\right)^{-\frac{\beta+1}{2\beta}}\right]
\end{eqnarray}
Then, the algebraic function (\ref{set39}) is characteristic of the singularity of type III.

\subsection{The Sudden singularity models without viscosity}
In this case, $-1<\beta<0$, and the corresponding scale factor reads
\begin{eqnarray}\label{set41}
a(t)=a_0e^{-(t_s-t)\left(H_s-\frac{h(t_s-t)^{-\beta}}{\beta-1}\right)}\,\,\,,
\end{eqnarray}
and the first derivative of the Hubble parameter remains the same as in (\ref{set27}), and Eq. (\ref{set13}) becomes
\begin{eqnarray}\label{set42}
2Tg_{TT}+g_T+\frac{\rho_0(1+\omega)(t_s-t)^{\beta+1}}{2\beta h a_0^{3(1+\omega)}}e^{3(1+\omega)(t_s-t)\left(H_s-\frac{h(t_s-t)^{-\beta}}{\beta-1}\right)}+1=0\,\,\,.
\end{eqnarray} 
By using (\ref{set14}), with $C$ substituted by $H_s$, as explained above and (\ref{set8}), one obtains
\begin{eqnarray}\label{set43}
\left(t_s-t\right)=\left[\sqrt{-\frac{T}{6h^2}}-\frac{H_s}{h}\right]^{-\frac{1}{\beta}}\,\,\,,
\end{eqnarray}
where $T_s$ denotes the torsion scalar at the singularity time, and may not be infinity. Using (\ref{set43}), Eq.(\ref{set42}) becomes
\begin{eqnarray}\label{set44}
2Tg_{TT}+g_T+\frac{\rho_0(1+\omega)}{2\beta h a_0^{3(1+\omega)}}\left[\sqrt{-\frac{T}{6h^2}}-\frac{H_s}{h}\right]^{-\frac{\beta+1}{\beta}}\times \nonumber \\e^{3(1+\omega)\left[\sqrt{-\frac{T}{6h^2}}-\frac{H_s}{h}\right]^{-\frac{1}{\beta}}\left(H_s-\frac{h}{\beta-1}\left[\sqrt{-\frac{T}{6h^2}}-\frac{H_s}{h}\right]\right)}+1=0\,\,\,.
\end{eqnarray}
This equation is not trivial and  cannot be solved analytically. However, as the singularity time is approached, with $-1<\beta<0$, it becomes
\begin{eqnarray}\label{set45}
2Tg_{TT}+g_T+\frac{\rho_0(1+\omega)}{2\beta h a_0^{3(1+\omega)}}\left[\sqrt{-\frac{T}{6h^2}}-\frac{H_s}{h}\right]^{-\frac{\beta+1}{\beta}}+1=0\,\,\,,
\end{eqnarray}
whose general solution reads
\begin{eqnarray}\label{set46}
g(T)&=&-T+\frac{2^{\frac{\beta+1}{\beta}}\beta h A}{\beta-1}\left[\sqrt{\frac{-4T}{9h^2}}-2\frac{H_s}{h}\right]\left[\sqrt{-6T}-6H_s\right]+2C_7\sqrt{-T}+C_8\,\,\,,
\end{eqnarray}
where $C_7$ and $C_8$ are integration constants, and $A$ previously defined in (\ref{set21}). The algebraic function $f(T)$ is then written as
\begin{eqnarray}\label{set47}
f(T)=\frac{2^{\frac{\beta+1}{\beta}}\beta h A}{\beta-1}\left[\sqrt{\frac{-4T}{9h^2}}-2\frac{H_s}{h}\right]\left[\sqrt{-6T}-6H_s\right]+2C_7\sqrt{-T}+C_8\,\,\,.
\end{eqnarray}
The constants $C_7$ and $C_8$ can be determined in the same way as in the previously cases. In this case, the first derivative of the torsion scalar with respect to the time, at initial time, is written as
\begin{eqnarray}\label{set48}
\left(\frac{dT}{dt}\right)_{t=t_0}=-12h\beta \left(-\frac{T_0}{6}\right)^{1/2}\left[\sqrt{-\frac{T_0}{6h^2}}-\frac{H_s}{h}\right]^{\frac{\beta+1}{\beta}}\,\,\,,
\end{eqnarray}
and (\ref{set24}) is valid. Then, we obtain the constants
\begin{eqnarray}
C_7= -\left[1+\frac{2^{\frac{5\beta+2}{2\beta}}\beta A}{\sqrt{3}(\beta-1)}\right]\sqrt{-T_0}+\frac{\beta A(2+\sqrt{6})2^{\frac{\beta+1}{\beta}}}{\beta}\,\,\,\,,\label{set49}\\
C_8=T_0-\frac{2^{\frac{\beta+1}{\beta}}\beta h A}{\beta-1}\left[\sqrt{\frac{-4T_0}{9h^2}}-2\frac{H_s}{h}\right]\left[\sqrt{-6T_0}-6H_s\right]-2C_7\sqrt{-T_0}\,\,.\label{set50}
\end{eqnarray}
Then, the algebraic function (\ref{set47}), with the constants $C_7$ and $C_8$ respectively in (\ref{set49}) and (\ref{set50}), is the $f(T)$ models that may produce the singularity of type II.

\subsection{The Big Brake models without viscosity}
Here, the scale factor is the same as in the case of the Sudden singularity, just that in this case, $\beta<-1$. Then, it can be easily observed that the differential equation (\ref{set45}) is also valid in this case. Consequently, the algebraic function (\ref{set47})  can lead to the Big Brake. The difference which appears here, with respect to the sudden singularity, is the values of the parameter $\beta$.

\section{Analysing the possible avoidance of the singularities in a viscous fluid}
Let us assume the fluid equation of state in the following form
\begin{eqnarray}
p=\omega \rho -3H\zeta(\rho)\,\,\,,\label{set54}
\end{eqnarray}
where $\zeta(\rho)$ is the bulk viscosity and in general depends on $\rho$. According to the thermodynamics grounds, the quantity $\zeta(\rho)$ has to be positive in order to guarantee the positive sign of the entropy change in an irreversible process. In this case, the stress-energy tensor $T_{\mu\nu}$ is written as 
\begin{eqnarray}
\mathcal{T}_{\mu\nu}=\rho u_{\mu}u_{\nu}-\left[\omega\rho-3H\zeta(\rho)\right]\left(g_{\mu\nu}-
u_{\mu}u_{\nu}\right)\,\,\,.\label{set55}
\end{eqnarray}
The equations of motion (\ref{set9}) and (\ref{set10}) can be written as 
\begin{eqnarray}
3H^3&=&\rho+\rho_T\,\,\,,\label{set56}\\
-2\dot{H}-3H^2&=&p+\rho_T\,\,\,,\label{set57}
\end{eqnarray}
where the modified gravity part is formally included into the modified energy density $\rho_T$ and the modified pressure $p_{T}$ as follows
\begin{eqnarray}
\rho_T&=&-\frac{1}{2}g-6H^2g_T\,\,\,,\label{set58}\\
p_{T}&=&\frac{1}{2}g+2\left(3H^2+\dot{H}\right)g_T-24\dot{H}H^2g_{TT}\,\,\,.\label{set59}
\end{eqnarray}
If we assume that the ordinary and the dark fluids of the universe do not interact, the  equation of continuity related to the ordinary viscous fluid reads
\begin{eqnarray}
\dot{\rho}+3H\rho\left(1+\omega\right)=9H^2\zeta(\rho)\,\,\,.\label{set60}
\end{eqnarray}
As we have seen in the section $3$, when the fluid does not possess viscosity, the four type of future finite-time singularities may appear. Now by introducing the viscosity, we analyse its effect near the singularities. Let us start with the constant viscosity case.
\subsection{The constant viscosity case: $\zeta(\rho)=\zeta_0$}
In order to analyse the effect of the viscosity on the Big Rip, Sudden, Big Freeze and Big Brake singular models, it is worth considering  the behaviour of the conservation law in Eq.~(\ref{set60}) near the singularities. By using Eq.~(\ref{set14}), one can write Eq.~(\ref{set60}) as
\begin{eqnarray}
\dot{\rho}+3\rho\left(1+\omega\right)\left[\left(t_s-t\right)^{-\beta}+C\right] \simeq 9\zeta_0h^2\left(t_s-t\right)^{-2\beta}+18\zeta_0Ch\left(t_s-t\right)^{-2\beta}+9\zeta_0C^2\,\,\,.\label{set61}
\end{eqnarray}
Solutions of Eq.~(\ref{set61}) can be found according to different values of the parameter $\beta$ as
\begin{eqnarray}
\rho_{vis}&=& \frac{9h^2\zeta_0}{(t_s-t)(1+3h+3h\omega)}\,\,\,, \quad \quad\quad\quad\quad\quad\quad\quad\quad\mbox{for}\quad \beta=1\,\,\,,\\
\rho_{vis} &\simeq& \frac{3h\zeta_0}{(1+\omega)(t_s-t)^\beta}\,\,\,,\quad \quad \quad\quad\quad\quad \quad \quad\quad\quad\quad \mbox{for}\quad \beta >1\,\,\,\,,\\
\rho_{vis} &\simeq& \frac{9\zeta_0h^2}{(2\beta-1)(t_s-t)^{2\beta-1}}\,\,\,,\quad \quad \quad \quad\quad\quad\quad\quad\quad\mbox{for}\quad 0<\beta<1\,\,\,,\\
\rho_{vis} &\simeq& \frac{9hH_s \zeta_0}{(\beta-1)(t_s-t)^{\beta-1}}+\frac{3H_s\zeta_0}{1+\omega}\,\,\,,\quad \quad \quad\quad\quad \mbox{for}\quad \beta <0\,\,,\,\,\, H_s\neq 0\,\,\,,
\end{eqnarray}
where we used $C=0$ for the Big Rip and the Big Freeze as previously discussed, since this constant is not relevant in these cases, while for the Sudden and the Big Brake, the constant $C$ is taken equal to the Hubble parameter $H_s$ at singularity time. Since $H_s$ plays a crucial role in these two cases, it has to be different from zero.\par
$\bullet$ For $\beta=1$, (Big rip), the energy density of the viscous fluid behaves as $(t_s-t)^{-1}$, while the Hubble parameter  behaves as $(t_s-t)^{-2}$. We see that the viscosity part of the energy density diverges more slowly than the Hubble parameter. Hence, in this case, the constant viscosity cannot avoid the occurrence of the Big Rip for $\beta=1$.\par
$\bullet$ For $\beta>1$, the Hubble parameter behaves as $(t_s-t)^{-2\beta}$, while the energy density from the viscous fluid behaves as $(t_s-t)^{-\beta}$, then, is less that the Hubble parameter. Hence, also  in this case, the Big Rip cannot be avoided by a fluid with constant viscosity.\par
$\bullet$ For $0<\beta<1$, the energy density from viscous fluid behaves as $(t_s-t)^{1-2\beta}$, while the Hubble parameter behaves as $(t_s-t)^{-2\beta}$. We see in this case that the energy density of the viscous fluid diverges less that the Hubble parameter. Hence, the Big Freeze cannot be avoided by constant viscosity.\par
$\bullet$ For $-1<\beta<0$, both the Hubble parameter and the energy density from the viscous fluid are finite. The task here is to consider the behaviour of the first derivative of the Hubble parameter $\dot{H}$ and the pressure of the viscous fluid. Note that as the Sudden singularity is approached, $\dot{H}$ behaves as $(t_s-t)^{-1-\beta}$, while the pressure $p_{vis}$ of the viscous fluid behaves as $(t_s-t)^{-2\beta}$. It appears that the pressure is finite while $\dot{H}$  diverges. Then, the Sudden singularity is robust against the constant viscosity. \par
$\bullet$ Since $\dot{H}$ diverges in this case of constant viscosity, the higher derivatives of $H$ also diverge, then the Big Brake cannot be avoided by constant viscosity.\par
We conclude that in $f(T)$ gravity and with the equation of state (\ref{set54}) for a fluid with constant viscosity, all the singularities are robust against the viscosity.
\subsection{The viscosity proportional to $\sqrt{-T}$}
From Eq.~(\ref{set8}) one can observe that this case is that in which the viscosity is proportional to the Hubble parameter. Let us then consider $\zeta=\tau\sqrt{-6T}/2=3\tau H$, where $\tau$ is a positive constant. The different solutions of (\ref{set60}) according to the values of the parameter $\beta$ read
\begin{eqnarray}
\rho_{vis}&=&\frac{27h^3\tau}{(t_s-t)^2(2+3h+3h\omega)}\,\,,\quad\quad\quad\quad \quad \quad\mbox{for} \quad \beta=1\,\,,\\
\rho_{vis}&\simeq& \frac{9h^2\tau}{(1+\omega)(t_s-t)^{2\beta}}\,\,, \quad\quad\quad\quad \quad \quad\quad\quad\quad\mbox{for} \quad \beta>1\,\,,\\
\rho_{vis}&\simeq&\frac{27h^3\tau}{(3\beta-1)(t_s-t)^{3\beta-1}}\,\,,\quad\quad\quad\quad\quad \quad\quad\;\;\mbox{for} \quad 0<\beta<1\,\,,\\
\rho_{vis}&\simeq&\frac{27h\tau H_s}{(\beta-1)(t_s-t)^{\beta-1}}+\frac{9\tau H_s}{1+\omega}\,\,\,,\quad\quad\quad\quad\;\;\mbox{for}\quad \beta<0\,\,, \quad H_s\neq 0\,\,.
\end{eqnarray}
Let us now discuss the effects of the viscosity near each type of singularity.\par
$\bullet$ For $\beta=1$, the energy density of the viscous fluid diverges like $H^2$. For small values of $\tau$, one can conclude that the Big Rip cannot be avoided. However, for large values of $\tau$ the Big Rip could be prevented.\par
$\bullet$ For $\beta>1$, $\rho_{vis}$ diverges like $H^2$. Thus, for small values of $\tau$, the Big Rip cannot be avoided, while for large values of $\tau$ it can be avoided.\par
$\bullet$ For $0<\beta<1$, $\rho_{vis}$ diverges like $(t_s-t)^{1-3\beta}$ while the $H^2$ diverges like $(t_s-t)^{-2\beta}$. For $0<\beta<1/3$, $\rho_{vis}$ is finite while for $1/3<\beta<1$, it diverges. However, in the case $\rho_{vis}$ diverges, it diverges  less than $H^2$. Thus the Big Freeze is robust against the viscosity.\par
$\bullet$ For $-1<\beta<0$, $\rho_{vis}$ tends to $9\tau H_s/(1+\omega)$, then, the pressure $p_{vis}$  tends to $9\omega\tau H_s/(1+\omega)$. At the same time the first derivative of the Hubble parameter, $\dot{H}$, diverges. Thus, for the small values of $\tau$, the viscous fluid cannot influence the Sudden singularity. However, for large values of $\tau$, the viscous pressure could dominate over the $\dot{H}$ end then, the Sudden singularity may be avoided\footnote{This is possible only if $\tau$ is sufficiently large for dominate over the divergence of $\dot{H}$.}.\par
$\bullet$ For $\beta<-1$, one can analyse the behaviour of the second derivative of the Hubble parameter, $\ddot{H}$, near the Big brake and compare it with the behaviour of the second derivative of $\rho_{vis}^{1/2}$. One can observe that $\ddot{H}$ diverges only if $-2<\beta<-1$. For the same interval, $\rho^{1/2}_{vis}$ also diverges as the singularity is approached, but less than $\ddot{H}$. Hence, the viscous fluid in this case cannot influence the Big Brake.
\subsection{More general case: $\zeta$ proportional to $(-T)^{n/2}$}
In this general case, we consider the bulk viscosity as 
\begin{eqnarray}
\zeta=\tau\left(\frac{-3T}{2}\right)^{\frac{n}{2}}=\tau(3H)^n\,\,\,,
\end{eqnarray}
where $n$ is a natural number, and $\tau$ a non null positive constant. Hence, the energy conservation leads to 
\begin{eqnarray}
\dot{\rho} +3H(\omega+1)=9H^2(3H)^n\tau\,\,\,,
\end{eqnarray} 
from which we obtain the behaviour of the energy density of the viscous fluid as
\begin{eqnarray}
\rho_{vis}&=&\frac{\tau(3h)^{n+2}}{n+1+3h(1+\omega)}(t_s-t)^{-n-1}\,\,\,,\quad\quad\quad\quad\quad\quad\quad\quad \mbox{for} \quad \beta=1\,\,\,\,,\\
\rho_{vis}&\simeq& \frac{(3h)\tau}{\omega+1}(t_s-t)^{-(n+1)\beta}\,\,\,,\quad\quad\quad\quad\quad\quad\quad\quad\quad\quad\quad\quad \mbox{for}\quad \beta>1\,\,\,,\\
\rho_{vis}&\simeq&\frac{(3h)^{n+2}\tau}{(2+n)\beta-1}(t_s-t)^{1-\beta(n+2)}\,\,\,,\quad\quad\quad\quad\quad\quad\quad\quad\;\;\mbox{for}\quad 0<\beta<1\,\,\,,\\
\rho_{vis}&\simeq&\frac{(3H_s)^{n+2}h\tau}{H_s(\beta-1)}(t_s-t)^{1-\beta}+\frac{(3H_s)^{n+1}\tau}{1+\omega}\,\,\,,\quad\quad\quad\quad\quad\mbox{for}\quad \beta<0\,\,\,.
\end{eqnarray}

$\bullet$ For $\beta=1$, $H^2$ diverges like $(t_s-t)^{-2}$, while $\rho_{vis}$ diverges like $(t_s-t)^{-1-n}$. Here the situation becomes discussable due to the presence on the parameter $n$. Hence, one observes that for $0<n<1$, as the singularity is approached, $H^2$ diverges more than  $\rho_{vis}$. Then, the Big Rip cannot be avoided by the viscous fluid. However, for $n>1$, the $\rho_{vis}$ diverges more than $H^2$ and then, the Big Rip can be avoided.\par
$\bullet$ When $\beta>1$, $\rho_{vis}$ diverges like $(t_s-t)^{-(n+1)\beta}$, while $H^2$ diverges like $(t_s-t)^{-2}$. Here, we see that for $n<-1+2/\beta$ (with $1<\beta<2$), $H^2$ diverges more than $\rho_{vis}$. Hence, the Big Rip is robust against the viscous fluid. But when $n>2\beta -1$, $\rho_{vis}$ dominates over $H^2$. Hence, the Big Rip may be avoided from the effect of the viscosity.\par
$\bullet$ For $0<\beta<1$, $H^2$ diverges like $(t_s-t)^{-2}$, while $\rho_{vis}$ diverges like $(t_s-t)^{1-\beta(n+2)}$. When $n<-2+3/\beta$ (with $1<\beta<3$), the energy density of the viscous fluid diverges less than $H^2$ and the Big Freeze cannot be avoided. But for $n>-2+3/\beta$, the energy density of the viscous fluid dominates over the background and the avoidance of the Big Freeze becomes possible.\par
$\bullet$ For $-1<\beta<0$, both the viscous energy density and $H^2$ are finites. In this case, it is necessary to  compare the behaviour of the pressure of the viscous fluid with the first derivative of the Hubble parameter, i.e. $\dot{H}$. Note that the pressure of the viscous fluid is also finite but depends strongly on the parameter $n$, that is, $p_{vis}=\omega\tau(3H_s)^{n+1}/(\omega+1)$, while $\dot{H}$ diverges like $(t_s-t)^{-\beta-1}$. Then, {\it a priori}, the Sudden singularity cannot be avoided by the viscous fluid. However for large values of $n$, $H_s$ and $\tau$, the pressure of the viscous fluid may dominate over $\dot{H}$ and the Sudden singularity could be avoided.\par 
$\bullet$ For $\beta<-1$, if the values of the parameters $n$, $H_s$ and $\tau$ are very large, the viscous fluid can influence the feature of the Big Brake, but not necessary avoid they.\par
As conclusion, we see that for a viscous fluid those  viscosity is proportional to $(-T)^{n/2}$, with the equation of state (\ref{set54}), the Big Rip and the Big Freeze can be eliminated for some values of $n$. However, in the case of the Sudden and the Big Brake, for large values of $n$, $H_s$ and $\tau$, the viscous fluid influences the singularities but does not necessary avoid they.

\section{Conclusion}
We considered in this work the modified teleparallel theory, known as $f(T)$ theory, where $T$ is the torsion scalar. In a specific way, the algebraic function $f(T)$ is taken as the teleparallel term $T$ plus the algebraic function $g(T)$. The equation of  motion of the theory is used and differential equation is established with the algebraic $g(T)$. The expression (\ref{set14}) is assumed for the Hubble parameter where the parameter $\beta$ plays an important role in specifying the type of singularity. Besides the parameter $\beta$ which is sufficient for characterizing the singularities, the Big Rip and the Big Freeze, we need to introduce  the parameter $H_s$, which substitutes the constant $C$,  for specifying the Sudden singularity and the Big Brake. Then, the differential equation is solved in each case and the corresponding algebraic function $f(T)$ which may lead to each type of singularity is obtained. All  this is done considering a fluid without the viscosity.\par
In order to probe the possible avoidance of the singularities, we introduce the bulk viscosity $\zeta(\rho)$ in three ways. The first case is when the viscosity is constant and then we observe that in general, the viscosity is inefficient against to the singularities. In  the second second where the viscosity is proportional to $\sqrt{-T}$, we see that for small values of the parameter $\tau$ the singularities are robust again the viscosity, while for large values of the this parameter, just the Big Rip and the Big Freeze may be cured  by the viscosity (the Sudden and the Big Brake remain robust in this case). In the third case where the viscosity is proportional to $(-T)^{n/2}$, we observe that independently of the values of $\tau$, the Big Rip and the Big Freeze may be avoided for some values of the parameter $n$. However, the Sudden and the Big Brake could be cured only for large values of $n$, $\tau$ and $H_s$.

\section*{Acknowledgements} M.J.S.Houndjo thanks IMSP for the Hospitality during the elaboration of this first version of this work and CNPq/FAPES for financial support. We also  thank Dr. Manuel E. Rodrigues and prof. S. D. Odintsov for useful suggestions.



\end{document}